\documentclass[aps,prd,preprintnumbers,superscriptaddress,nofootinbib,floatfix,twocolumn,notitlepage]{revtex4-2}
\usepackage{graphicx}  
\usepackage{dcolumn}   
\usepackage{bm}        
\usepackage{epsfig,amsmath,amssymb,verbatim,mathrsfs,array,layout,textcomp,latexsym,slashed}
\usepackage{xcolor}

\usepackage{esvect}
\usepackage{multirow}

\usepackage[left]{lineno}
\usepackage{blindtext}
\pdfoutput=1
\usepackage[T1]{fontenc}
\usepackage{amsfonts}
\usepackage{braket}
\usepackage{cancel}
\usepackage{pstricks}
\usepackage{feynmf}
\usepackage[normalem]{ulem}
\usepackage{epstopdf}
\usepackage{soul}
\usepackage{cancel}
\def\beq#1\eeq{\begin{align}#1\end{align}}

\definecolor{ao(english)}{rgb}{0.0, 0.5, 0.0}

\usepackage{braket}
\usepackage{cancel}
\usepackage{setspace}
\usepackage{pstricks}
\usepackage{url}
\usepackage{xcolor}
\usepackage{float}
\usepackage{mathtools}
\usepackage{multirow}
\usepackage{booktabs}
\usepackage{enumitem}

\RequirePackage{xspace}
\def\Bbar    {\kern 0.18em\overline{\kern -0.18em B}{}\xspace}

\newcommand{\ie}{{\em i.e.}}
\usepackage{tabularx}
\newcolumntype{Y}{>{\centering\arraybackslash}X} 
\usepackage{booktabs} 
\usepackage{colortbl} 
\RequirePackage{xspace}
\def\Bbar    {\kern 0.18em\overline{\kern -0.18em B}{}\xspace}

\newcommand{\eq}[1]{Eq.~(\ref{#1})}

\definecolor{BlueViolet}{rgb}{0.2, 0.00, 0.7}
\definecolor{Blue}{rgb}{0.15, 0.00, 0.9}
\definecolor{lightblue}{rgb}{0.284602,0.317763,0.963947}
\definecolor{kitgreen}{rgb}{0, 
0.58823 
, 0.50980 
}
\definecolor{rossoferrari}{HTML}{D9073D}
\definecolor{Pink}{HTML}{FF686E}
\usepackage[colorlinks=true, 
linkcolor=lightblue,
citecolor=lightblue,
urlcolor=kitgreen]{hyperref} 
\usepackage[hyphenbreaks]{breakurl} 

\begin{document}

\title{MeV Sterile Neutrino in light of the Cabibbo-Angle Anomaly}
\preprint{KEK--TH--2544}

\author{Teppei Kitahara}
\email{teppeik@itp.ac.cn}
\affiliation{CAS Key Laboratory of Theoretical Physics, Institute of Theoretical Physics, Chinese Academy of Sciences, Beijing 100190, China}
\affiliation{Kobayashi-Maskawa Institute for the Origin of Particles and the Universe, 
Nagoya University,  Nagoya 464--8602, Japan}

\author{Kohsaku Tobioka}
\email{ktobioka@fsu.edu}
\affiliation{Department of Physics, Florida State University, Tallahassee, FL 32306, USA}
\affiliation{KEK Theory Center, IPNS, KEK, Tsukuba  305--0801, Japan}

\begin{abstract}
A modified neutrino sector could imprint a signature on precision measurements of the quark sector because many such measurements rely on the semi-leptonic decays of the charged currents. 
Currently, global fits of the determinations of the Cabibbo-Kobayashi-Maskawa (CKM) matrix elements point to 
a $3\sigma$-level deficit 
in the first-row CKM unitarity test,
commonly referred to as the Cabibbo-angle anomaly. 
We find that a MeV sterile neutrino that mixes with the electron-type neutrino increases the extracted $|V_{ud}|$, accommodating the Cabibbo-angle anomaly.
This MeV sterile neutrino affects the superallowed nuclear $\beta$ decays and neutron decay, but it barely modifies the other measurements of the CKM elements. While various constraints may apply to such a sterile neutrino, we present viable scenarios within an extension of the inverse seesaw model. 
\end{abstract}

\maketitle

\section{Introduction}
While precision measurements have largely confirmed the predictions of the standard model (SM), there are now unignorable deviations. Notably, certain observables, such as the lepton flavor universality in meson decays \cite{HFLAV:2022pwe}, have begun to diverge from SM predictions. Against this backdrop, the neutrino sector, often considered a relatively unexplored domain in direct determinations, offers intriguing possibilities. Many of these anomalies seemingly in the quark sector could be actually due to new physics (NP) of the neutrino sector, given that the relevant measurements heavily rely on the semi-leptonic decays involving neutrinos.

In this paper, we consider, as one such example, the unitarity relation of the Cabibbo-Kobayashi-Maskawa (CKM) matrix $V V^\dag ={\boldsymbol I}_3$ \cite{Cabibbo:1963yz, Kobayashi:1973fv}. 
A deficit has been observed in the first-row CKM unitarity test \cite{Seng:2018yzq,Belfatto:2019swo,Grossman:2019bzp,Coutinho:2019aiy,Cheung:2020vqm,Seng:2020wjq,Crivellin:2020ebi, Kirk:2020wdk, Crivellin:2021njn,Belfatto:2021jhf,Bryman:2021teu,Cirigliano:2022yyo,Crivellin:2022rhw,Belfatto:2023tbv}.
According to  a 
recent review in Ref.~\cite{Crivellin:2022rhw}, 
 we obtain
\begin{align}
(V V^\dag)_{11} \equiv |V_{ud}|^2 +|V_{us}|^2 + |V_{ub}|^2 =1
+ \Delta_{\rm CKM}^{\rm global} \,, 
 \end{align}
with 
\begin{align}
 \!\! \Delta_{\rm CKM}^{\rm global}   =
    \left\{
    \begin{array}{lll}
  \!\!  -1.51(53)\times 10^{-3}
    ~ \text{(w/\,bottle\,UCN\,best)}
    \,,\\
 \!\!   -2.34(62)\times 10^{-3}
     ~\text{(w/\,in-beam\,best)}
     \,,
    \end{array}
    \right. \!\!
    \label{eq:dletaCKM}
\end{align}
which deviate from the unitarity relation ($\Delta_{\rm CKM}^{\rm global} =0$) 
at a significance of $-2.8\sigma$ and $-3.8\sigma$, respectively. 
Here, all available data from the kaon, pion, tau lepton, hyperon decays, and various types of $\beta$ decays are used in the global fit \cite{Crivellin:2022rhw}, and all the decays involve neutrinos.
The difference in \eq{eq:dletaCKM} comes from the two different input data of the neutron decays:
\beq
|V_{ud,n}|= 0.974\,13(43)\quad \text{(bottle\,UCN\,best)}\,,
\label{eq:bottle}
\eeq
based on the single most precise bottle ultracold neutron (UCN) lifetime data $\tau_n^{\rm bottle} =877.75(36)$\,sec~\cite{UCNt:2021pcg},
while
\beq
|V_{ud,n}|= 0.968\,66(131)\quad \text{(in-beam\,best)}\,,
\label{eq:inbeam}
\eeq
  based on the in-beam one $\tau_n^{\rm beam}=887.7(2.2)$\,sec
  \cite{Yue:2013qrc}.
Here, the single most precise data of the nucleon isovector axial charge $g_A/g_V$\,\cite{Markisch:2018ndu} and an updated radiative correction $\Delta_R$  
\cite{Cirigliano:2022yyo} are used in both cases.
It is found that the beam neutron data provides a little bit larger significance in the deficit.
Note that as discussed in the following, $|V_{ud}|$ is mostly determined by the data of the superallowed  nuclear $\beta$ decays.

Since the size of $|V_{ub}|^2$ is significantly smaller than uncertainties of the other components  \cite{Charles:2004jd,UTfit:2022hsi},
the deficit is essentially inherent in the two-generation Cabibbo angle. 
Hence, it is referred to as the {\em Cabibbo angle anomaly}~(CAA) \cite{Grossman:2019bzp,Coutinho:2019aiy}. 
On the other hand, the ordinary CKM unitarity triangle  (in $B$-meson decays) corresponds to $(V^\dag V)_{31}=0$, which is currently consistent with the SM prediction
within the experimental errors \cite{UTfit:2022hsi}.
 The representative measurements specific to the $(V V^\dag)_{11}$ unitarity test are:
 the superallowed nuclear $\beta$ decay and the neutron decay for $|V_{ud}|$;
the kaon decays such as $K \to \pi \ell \nu$, $K \to \mu \nu$, 
and tau hadronic decays for $|V_{us}|$ and their ratio $|V_{us}/V_{ud}|$.
The recent global fit of them and assessment of the deficit of $(V V^\dag)_{11}$ have been performed in Refs.~\cite{Cirigliano:2022yyo,Crivellin:2022rhw}.

There have been several attempts to resolve this anomaly by TeV-scale  new physics models \cite{Belfatto:2019swo,Coutinho:2019aiy,Cheung:2020vqm,Crivellin:2020lzu,Crivellin:2020ebi,Kirk:2020wdk,Crivellin:2021njn,Belfatto:2021jhf,Bryman:2021teu,Cirigliano:2022yyo,Crivellin:2022rhw,Belfatto:2023tbv,Endo:2020tkb,Capdevila:2020rrl,Li:2020wxi,Crivellin:2020oup,Crivellin:2020klg,Felkl:2021qdn,Branco:2021vhs,Marzocca:2021azj,Buras:2021btx,Crivellin:2021bkd,Cirigliano:2021yto,Blennow:2022yfm}. In this article, we develop deeper the extensions of neutrino sector concerning the anomaly. Our findings uniquely suggest that the CAA might be pointing to an intermediate mass scale for new particles, specifically, a sterile neutrino at the MeV scale that mixes with the electron-type neutrino~\cite{Bryman:2019bjg,Kitahara:2023kmh}. This scenario could arise because all the relevant precision measurements use (semi-)leptonic decays and involve neutrinos, considering that theoretical and experimental uncertainties of these decays are well managed. As we will show, the MeV sterile neutrino could be the underlying cause of the anomaly seemingly in the quark sector quantities. It is worth noting that several studies have indicated that the CAA is not easily resolved by heavy or massless sterile neutrinos~\cite{Coutinho:2019aiy,Blennow:2022yfm, Li:2020wxi}.

In the next section, we delve into the contributions of sterile neutrinos to the observables for the CAA. 
In Sec.~\ref{sec:constraint},
we summarize and discuss the current constraints on the favored parameter region, and conclude in Sec.~\ref{sec:conclusion}.


\section{Sterile Neutrino in light of CAA}
\label{sec:review}
This section outlines how a sterile neutrino might resolve the CAA.
We initially examine two simple limits: very heavy and massless sterile neutrinos and show that they cannot accommodate the CAA. 
On the other hand, a MeV-scale sterile neutrino presents a potential solution by influencing the $|V_{ud}|$ determinations. To ascertain this solution, we conduct global fits in search of the optimal parameter space. 
(Discussions on experimental constraints and underlying models will be presented in Sec.~\ref{sec:constraint}.)

\begin{figure}[t]
\begin{center}
\includegraphics[width=0.49\textwidth]{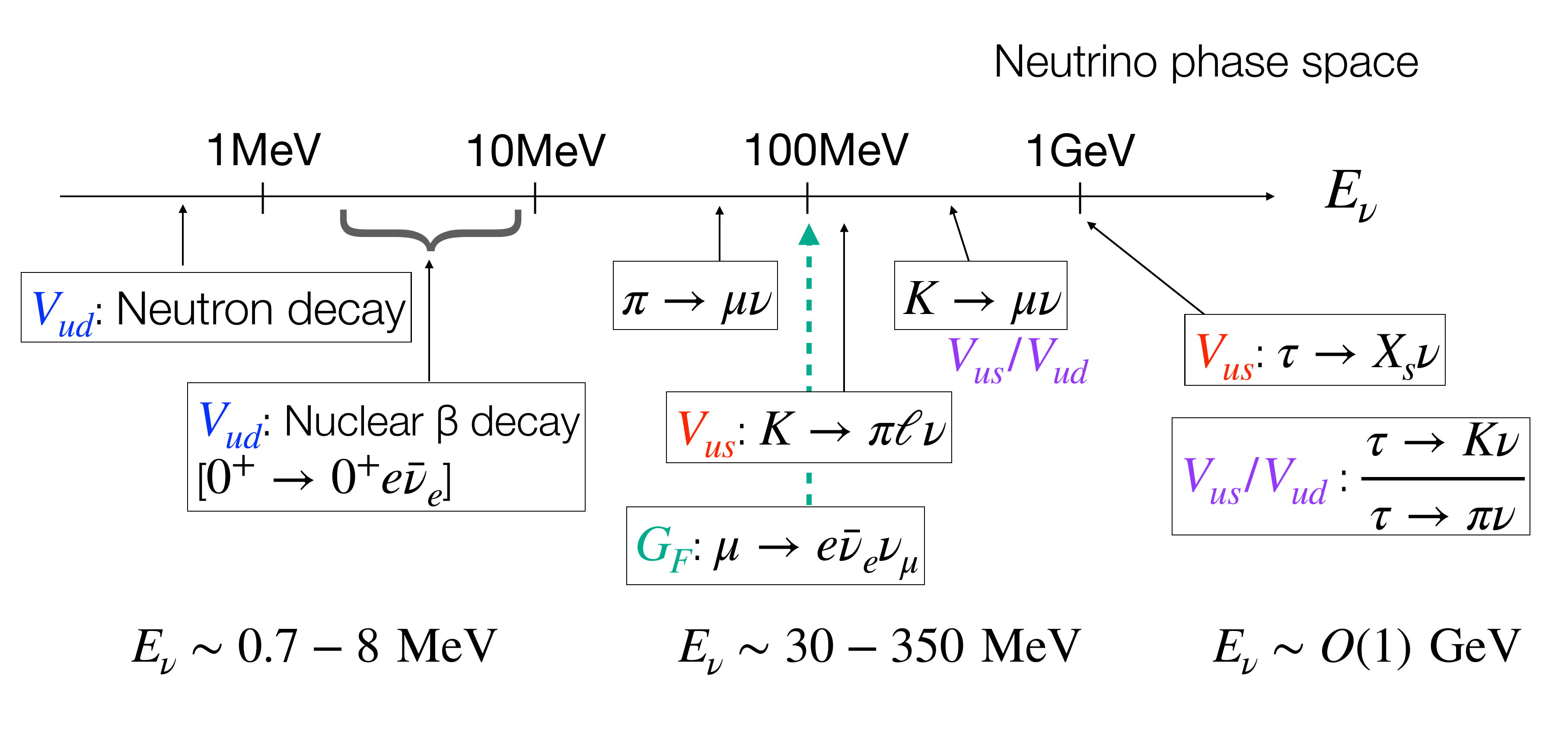}
\vspace{-10pt}
\caption{\label{fig:Enu}
Neutrino energy $E_\nu$ of each channel relevant to the first-row CKM unitarity test is summarized. The sterile neutrino channel is open when $m_{\nu 4} < E_{\nu}$.
}
\end{center}
\end{figure}

Typically, a sterile neutrino's characteristics in most models can be effectively captured by its mass, $m_{\nu 4}$, and its mixing angles with the SM neutrinos, $U_{\ell 4}$.
Below the electroweak scale, the SM neutrinos from the weak doublets, $\nu_\ell$, split into two pieces in the mass eigenbasis, 
\begin{align}
\nu_\ell \simeq \cos U_{\ell 4} \nu_{\ell}^\prime + \sin U_{\ell 4}  \nu_{4} \quad \text{for~}\ell=e,\mu,\tau\,,
\label{eq:mixing}
\end{align}
where $\nu_{\ell}^\prime$ are the SM-like neutrinos, often called active neutrinos.
For the active neutrinos, the mixing ($\cos U_{\ell 4} \lesssim 1$) results in the coupling reduction of the weak interaction, and the deficit gives the sterile neutrino a feeble coupling ($\sin U_{\ell 4} \ll 1$) to the SM.   
These couplings alter observables primarily governed by the weak interaction, \ie, all the measurements relevant to the CAA are potentially affected, see Fig.~\ref{fig:Enu}. 
The sign of the modification depends on the mass scale of the sterile neutrino, pinpointing a specific scale.

\subsection{Sterile neutrino above GeV}

If the sterile neutrino mass exceeds $\mathcal O$(1)~GeV, it would not be kinematically permissible in the relevant measurements. In this case, an important modification occurs in the Fermi constant measurement via the muon decay. 
The observed value, $G_F^{\rm obs}$, deviates from the true value, $G_F$,  as described by
\begin{align}
G_F^{\rm obs} =G_F \cos U_{e 4}  \cos U_{\mu 4}\,. \label{eq:GF}
\end{align}
This relationship is pivotal in all the measurements employed in the CKM determinations.  
The CKM element $V_{ud}$ is obtained  by the superallowed nuclear $\beta$ decay and neutron decay associated with $\nu'_e$, and we have $V_{ud}^{\rm obs}  G_F^{\rm obs} =V_{ud} G_F \cos U_{e 4}$  which leads to
\begin{align}
V_{ud}^{\rm obs} &=V_{ud} /\cos U_{\mu 4}\,.  \label{eq:Vud1}
\end{align}
Other elements $V_{us}$ and its ratio $V_{us}/V_{ud}$ are derived from the measurements involving $\nu'_\mu$.\footnote{$V_{us}$ is also determined from the electron mode, $K\to \pi e \nu_e$ (with better precision than the muon mode), which leads to $V_{us}^{\rm obs} =V_{us} /\cos U_{\mu 4}$. But,  combining Eq.~\eqref{eq:Vud1}, $\Delta_{\rm CKM}>0$ is predicted in either way.} Analogous to Eq.~\eqref{eq:Vud1}, the relationships between the observed and true values are given by
\begin{align}
V_{us}^{\rm obs} &=V_{us} /\cos U_{e 4} \,, \label{eq:Vus1}  \\
(V_{us}/V_{ud})^{\rm obs} &=V_{us}/V_{ud}\,. \label{eq:Vus/Vud1}
\end{align}

Now one can check the unitarity relation, for example, combining Eqs.~\eqref{eq:Vud1} with \eqref{eq:Vus1}, 
\begin{align}
1+\Delta_{\rm CKM} = |V_{ud}^{\rm obs}|^2+|V_{us}^{\rm obs}|^2 =\frac{|V_{ud}|^2}{\cos^2 U_{\mu 4}}+\frac{|V_{us}|^2}{\cos^2 U_{e 4}}\,. 
\end{align}
Given the inherent CKM unitarity, represented by $|V_{ud}|^2+|V_{us}|^2\simeq 1$, it follows $\Delta_{\rm CKM} >0$, 
which is in contrast to the experimental fits presented in Eq.~\eqref{eq:dletaCKM}.\footnote{The $m_W$ anomaly seen at the CDF experiment \cite{CDF:2022hxs} can be resolved in this framework, because the modification goes in the right direction.  Modification of $G_F$ is essential in this case. See also Ref.~\cite{Blennow:2022yfm}.} Similarly, combining Eqs.~\eqref{eq:Vud1} and \eqref{eq:Vus/Vud1} draws the same conclusion.

\subsection{Light sterile neutrino}

Given that a heavy sterile neutrino cannot address the anomaly, it is instructive to examine the opposite scenario where the sterile neutrino mass is significantly smaller than $E_\nu$, the maximum energy of neutrino in the relevant processes.
Intriguingly, under these conditions, experimental observations align with the SM predictions.
This is because the processes involving $\nu_4$ are kinematically allowed and must be summed incoherently. To illustrate, consider the muon decay where we account for four processes, 
\begin{align}
&\sum_{\nu_i=\nu'_\mu,\nu_4 ,~ \bar\nu_j=\bar\nu'_e,\bar\nu_4}
\Gamma(\mu\to e \nu_i \bar\nu_j)
\nonumber\\\
&\qquad \simeq (\cos^2 U_{e 4}+\sin^2 U_{e 4})(\cos^2 U_{\mu 4}+\sin^2 U_{\mu 4}) \Gamma_{\rm SM}\nonumber \\
&\qquad =\Gamma_{\rm SM}\,.
\label{eq:light_sterile}
\end{align}
Similarly, other processes reproduce the SM predictions. Hence, $\Delta_{\rm CKM}$ is expected to be zero,  which is again incompatible with the current data.

\subsection{MeV sterile neutrino interacting with electron}
The above discussion tells the CAA may suggest an intermediate mass scale of the sterile neutrino. If the mass is in the MeV scale, the coupling reduction of the weak interaction remains in the neutron/nuclear decays, while other observables, especially the Fermi constant, stay almost the same as in the SM as seen in the Eq.~\eqref{eq:light_sterile}. Consequently, only $V_{ud}$ is modified, 
\begin{align}
V_{ud}^{\rm obs} = V_{ud} \cos U_{e 4} \,.\label{eq:vudobs1}
\end{align}
The other quantities, $G_F^{\rm obs}, V_{us}^{\rm obs}$, and $(V_{us}/V_{ud})^{\rm obs}$, are the same as the SM ones because the corresponding $E_\nu$ is much larger than MeV. This realizes the experimentally favored value,
\begin{align}
1+\Delta_{\rm CKM} = |V_{ud}|^2 \cos^2 U_{e 4} +|V_{us}|^2 <1 \,.  
\end{align}
From the size of the anomaly $\Delta_{\rm CKM} \approx 10^{-3}$,
we can infer that the favored mixing-angle-squared is $U^2_{e 4}\approx 10^{-3}$, and $U^2_{\mu 4}$ is not necessary.\footnote{In the interest of the $m_W$ anomaly,
this parameter space is not necessarily favored, but it can be easily fixed if another heavy sterile neutrino ($\nu_5$) affects $G_F$ by $U_{\mu 5}$. Then, $V_{ud}$ is modified as
$V_{ud}^{\rm obs}\simeq V_{ud}(1-U_{e 4}^2/2+U_{\mu 5}^2/2)$.
}

 The sterile neutrino mass can be in the same order of $E_\nu$ of the neutron/nuclear decays relevant to $V_{ud}$ measurements.
In the following, we evaluate the sterile mass dependence in the neutron and nuclear decays, and we focus on the mixing with the electron neutrino.
   
\subsubsection{Superallowed nuclear \texorpdfstring{$\beta$}{beta} decay}
The measurements of $V_{ud}$ in nuclear physics have been conducted through so-called superallowed $0^+\to 0^+$ nuclear $\beta$ decay, 
and the latest survey was given by Ref.~\cite{Hardy:2020qwl}. A heavy nucleus of $J^P =0^+$ decay to another nucleus of $0^+$ with a significant wave function overlap ({\em superallowed}), emitting  $e^+ \nu_e$ ($\beta^+$ decay).
Also, 
only the vector current of the quark weak interaction can contribute  (Fermi decay). 
There are 15 transitions utilized to determine $|V_{ud}|$ \cite{Hardy:2020qwl}, and the released energy in these transitions varies according to the specific process. Note that the  determination of $V_{ud}$ is significantly dominated by a single transition, ${}^{26m}\text{Al}\to{}^{26}\text{Mg}$  \cite{Brindhaban:1994zz,Eronen:2006if}. 
While a sterile neutrino might be absent in some processes, it could emerge in others with larger released energy, and the effect is suppressed by the phase space as well as the small mixing angle $\sin^2 U_{e4}$.

Let us explore how $V_{ud}$ is measured and the impact of the massive sterile neutrino.  
The decay width for both the active and sterile neutrinos is given by 
\begin{align}
\!\!\!\! \frac{d\Gamma_{0^+}}{dE_e} =\left(\frac{ \sqrt{E_e^{2}-m_e^{2}}\sqrt{E_\nu^{2}-m_\nu^{2}}E_e E_\nu}{16\pi^3} \right) \frac{|{\cal M}|^2}{E_eE_\nu M M'}\,,
\end{align}
where the parenthesis comes from the phase space calculation. The outgoing neutrino (positron) energy is 
$E_{\nu(e)}$, and $M(M')$ is the mass of the parent (daughter) nucleus. 
$\delta M$ denotes the nucleus mass difference, which is $\delta M \equiv M - M'=  Q_{\rm EC}-m_e$ neglecting the electron binding energy of $\mathcal{O}(10)\,\text{eV}$~\cite{NIST_ASD}.  The 
 electron-capture (EC) $Q$-value $Q_{\rm EC}$ is the experimentally measured quantity summarized in Table I of Ref.~\cite{Hardy:2020qwl}. 
For example, $Q_{\rm EC}({{}^{26m}{\rm Al}\to {}^{26}{\rm Mg} })=4.23$\,MeV. 
The nucleus recoil energy $E_{\rm recoil}$ can be ignored because it is much smaller than the typical electron and nuclear energy. Thus  the energy conservation is approximately held as $\delta M\simeq E_\nu + E_e $, leading to $m_e\leq E_e  \leq \delta M -m_\nu$. This approximation greatly simplifies the calculation, and  is valid until the precision becomes as good as $\delta M /M\sim 10^{-4}$.   
With this accuracy, the matrix element is given by   
\begin{align}
|{\cal M}|^2\simeq
C E_eE_\nu M^2 G_F^2 |V_{ud}|^2 \,, \label{eq:ME}
\end{align}
where $C$ is a numerical coefficient.
Normally, one can extract the combination of $G_F^2 |V_{ud}|^2$ based on the ${\cal F} t$ values of each nuclear $\beta$ decay, which will be described later, under the assumption that the neutrinos are massless.

\begin{figure}[t]
\begin{center}
\includegraphics[width=0.48\textwidth]{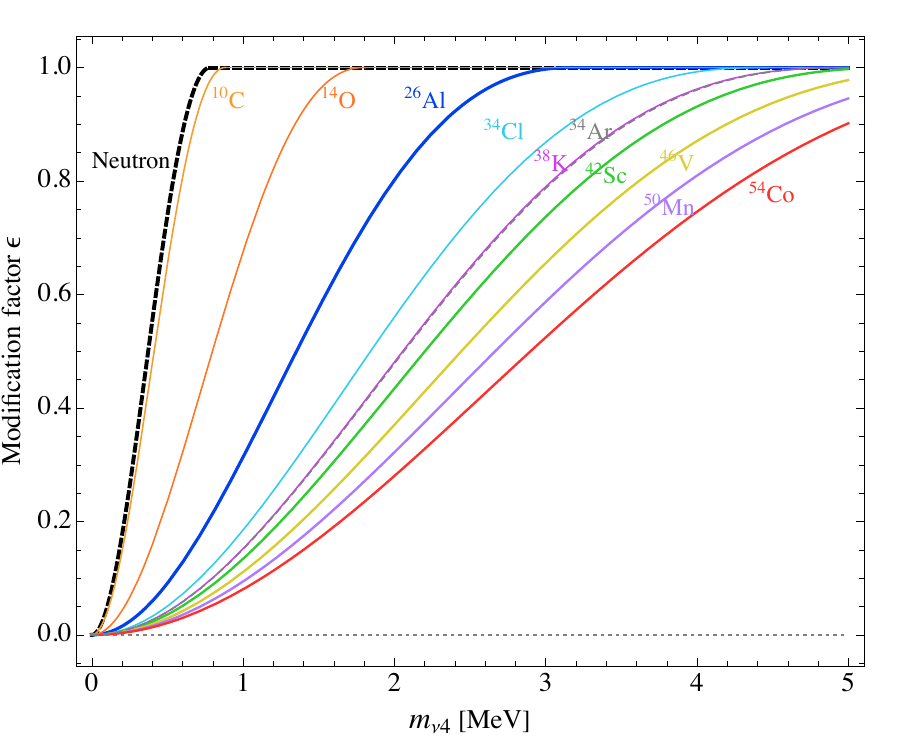}
\caption{
The phase space modification factors $\epsilon(m_{\nu 4}, \delta M)$ are shown for the ten most accurate superallowed decays (solid lines and a dotted line) and the neutron 
 decay (black dashed line). 
\label{fig:phasespace}}
\end{center}
\end{figure}

We present a simple formalism to account for the modification due to the massive sterile neutrino.
Given the mass of neutrino barely modifies the matrix element of Eq.~\eqref{eq:ME},  
the massive neutrino effect appears through the phase space. 
We define the modified phase space integral as 
\begin{align}
\!\!\! I\left(m_\nu, \delta M\right) \equiv \int_{m_e}^{\delta M-m_\nu}\!\!\!\!\!\! dE_e \sqrt{E_e^{2}-m_e^{2}}\sqrt{E_\nu^{2} -m_\nu^{2}} E_e E_\nu\,, 
\end{align}
where $I(0, \delta M)$ corresponds to the SM case (maximizes for a given $\delta M$), and    the kinematically forbidden neutrino yields $I(m_\nu>\delta M-m_e, \delta M)=0$.  
A single decay mode is given by the sum of the active and sterile neutrino channels. To extract the modification factor depending on the sterile neutrino mass and mixing angle, we normalize the decay width  by the one with $m_{\nu 4}=0$, 
\begin{align}
\frac{\Gamma_{0^+}}{\Gamma_{{0^+},m_{\nu 4}=0}} &= \cos^2 U_{e 4}   +  \frac{I(m_{\nu 4}, \delta M)}{I(0, \delta M)}\sin^2 U_{e4} 
\label{eq:betaPhaseSpace2}
\\
&=
1- \epsilon(m_{\nu 4}, \delta M)\sin^2 U_{e4} \,, \label{eq:betaPhaseSpace}
\end{align}
with
\begin{align}
\epsilon\left(m_{\nu 4}, \delta M\right)\equiv 1-\frac{I(m_{\nu 4}, \delta M)}{I(0, \delta M)}\,,
\end{align}
where $m_{\nu 4}$ represents the sterile neutrino mass. 
We found that   Eq.~\eqref{eq:betaPhaseSpace2} is consistent with a result of Ref.~\cite{Bryman:2019bjg}.
In Fig.~\ref{fig:phasespace},
the phase space modification factor $\epsilon(m_{\nu 4}, \delta M)$ is plotted for each superallowed channel. 
There, the charge repulsion or attraction effect is included in accordance with Ref.~\cite{Kleesiek:2018mel}, which has little impact.
The measured $|V_{ud,0^+}^{\rm obs}|$ through the nuclear $\beta$ decay is different from the value in the presence of the MeV sterile neutrino, and, therefore, we obtain 
\begin{align}
    \Gamma_{0^+}^{\rm obs}\propto  |V_{ud,0^+}^{\rm obs}|^2
    = \left[
    1- \epsilon(m_{\nu 4}, \delta M)\sin^2 U_{e4}\right] |V_{ud,0^+}|^2
\end{align}
by using Eq.~\eqref{eq:betaPhaseSpace}.
This reproduces the concise expression of Eq.~\eqref{eq:vudobs1} when $m_{\nu 4}>\delta M-m_e$, corresponding to $\epsilon=1$.
Thus, we expect $U_{e 4}^2\approx\Delta_{\rm CKM}^{\rm global}\approx 10^{-3}$ to accommodate the CAA.

\subsubsection{Neutron decay}
The neutron decay is currently the second-best probe of $V_{ud}$. 
The neutron lifetime ($\tau_n$) measurements and the theoretical calculations are put together to obtain $V_{ud,n}$. 
It is known that there are two different measurements for the neutron lifetime, so-called neutron decays in the bottle or in the beam, and these data indicate $\approx 4\sigma$ discrepancy
\cite{Czarnecki:2018okw}.
Although they are controversial, 
both results imply the violation of CKM unitarity, see details in the Introduction.
In our fit, we treat two methods separately and combine it with other measurements. 
The sensitivity for  $V_{ud}$  is still dominated by the superallowed nuclear decay, but the favored sterile neutrino mass changes a little depending on the method of neutron lifetime measurements.

The effect of massive sterile neutrino is very similar to the case of superallowed nuclear decay.  The neutron decay width is given by
\begin{align}
\frac{\Gamma_{n}}{\Gamma_{{n},m_{\nu 4}=0}} 
& = 1- \epsilon\left(m_{\nu 4}, \delta M_{np}\right) \sin^2 U_{e4}
\,,  \label{eq:betaPhaseSpaceN}
\end{align}
where $\delta M_{np}$ is the mass difference between a neutron and a proton.
Hence, the observed $V_{ud,n}^{\rm obs}$ can be smaller than the true  value, 
\begin{align}
    |V^{\rm obs}_{ud,n}|^2 =\left[
    1-\epsilon\left(m_{\nu 4}, \delta M_{np}\right) \sin^2 U_{e 4}\right]  |V_{ud,n}|^2 \,. 
\end{align}
The phase space modification factor $\epsilon\left(m_{\nu 4}, \delta M_{np}\right)$ is shown in Fig.~\ref{fig:phasespace}.

As we already addressed,
the neutron lifetime measurements are mutually inconsistent between the bottle UCN and the beam methods.
In this article, we treat them separately rather than combining. 

\begin{figure*}[t]
\begin{center}
\includegraphics[width=0.45\textwidth]{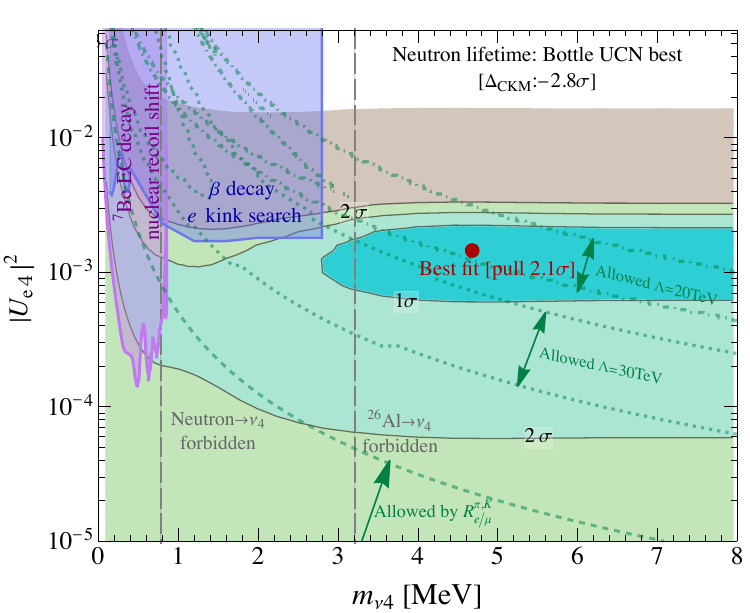}~~
\includegraphics[width=0.45\textwidth]{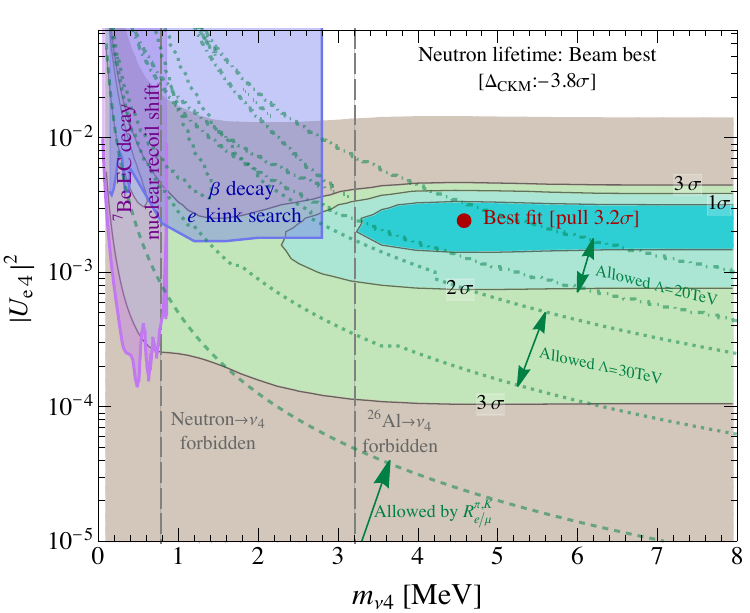}
\caption{Favored parameter regions are shown.
The red points represent the best-fit point with the pulls from the SM hypothesis.
For the neutron decay data, 
the most precise neutron lifetime results from the bottle UCN and the beam measurements are used in the left and right panels, respectively. 
The blue and purple shaded regions are excluded by the nuclear $\beta$-decay kink searches  \cite{Schreckenbach:1983cg,Deutsch:1990ut,Derbin:2018dbu,Bolton:2019pcu} and the EC-decay search  \cite{Friedrich:2020nze}, respectively.
The regions above green dashed lines can be constrained by $\pi^+ \to e^+ \nu$ measurements \cite{PiENu:2015seu,Bryman:2019ssi,Bryman:2019bjg}, and the regions between dotted or dash-dotted green lines are allowed with the dimension-six operator. See more details in the text.
\label{fig:neutrinofit}}
\end{center}
\end{figure*}

\subsubsection{Statistical combinations}

Basically we construct $\chi^2$ functions of $m_{\nu 4},~ \sin^2U_{e 4}\simeq U_{e 4}^2,~ |V_{ud}|$, and $|V_{us}|$, 
and impose the unitarity at the truth level, 
$|V_{ud}|^2 +|V_{us}|^2 =1$, safely dropping $|V_{ub}|^2 \simeq 1.4 \times 10^{-5}$ for the uncertainty of $\Delta_{\rm CKM}$ \cite{Charles:2004jd,UTfit:2022hsi}. 
Also, we do not include the pion $\beta$ decay $\pi^+ \to \pi^0 e^+ \nu$, which can independently measure $V_{ud}$ without nuclear corrections, in our analysis, because its current sensitivity is approximately ten times weaker than that of superallowed $\beta$ decay \cite{Cirigliano:2022yyo}.
In this subsection, we omit the absolute value notation for simplicity. 

We combine the $V_{ud}$ measurements which depend on the sterile neutrino mass less than $10\,$MeV and the mixing, 
with $V_{us}$ and $V_{us}/V_{ud}$ measurements which have negligible shifts from the sterile neutrino, since $U_{e 4}^2(m_{\nu4}/E_\nu)\lesssim 10^{-4}$. 
We adopt the following fitted values: 
 $V_{us}$ from the semi-leptonic kaon decays ($K_{\ell 3}$), inclusive-hadronic $\tau$ decays, and hyperon semi-leptonic decays,
 and $V_{us}/V_{ud}$ from the leptonic kaon-decay rate over the pion one ($K_{\mu 2}/\pi_{\mu 2}$) \cite{Crivellin:2022rhw}
\begin{align}
 V_{us, {\rm fit}} &=  0.223\,14 (51)\,,\\
 \left(V_{us}/V_{ud}\right)_{\rm fit} &= 0.231\,08 (51)\,.
\end{align}
Then, we construct two $\chi^2$ functions, 
\begin{align}
&\chi^2_{us} \left(V_{ud}\right)
= \frac{\left(\sqrt{1-V_{ud}^2} -V_{us,\rm fit } \right)^2}{\left(\delta V_{us}\right)^2}\,, 
\\
&\chi^2_{us/ud} \left(V_{ud}\right)
= \frac{\left[\sqrt{1-V_{ud}^2}/V_{ud} - (V_{us}/V_{ud})_{\rm fit } \right]^2}{\left[\delta (V_{us}/V_{ud})\right]^2}\,, 
\end{align}
where $\delta V_{us}$ represents the uncertainty of $V_{us,\rm fit }$, and so does $\delta (V_{us}/V_{ud})$.

For the $V_{ud}$ determination through the neutron lifetime, we have to consider the effect of massive sterile neutrino, and then we have 
\begin{align}
\!\!\!\!\!\! \chi^2_{n} (m_{\nu 4}, U_{e 4}^2, V_{ud})= 
\frac{\left[V_{ud}\sqrt{1- \epsilon(m_{\nu 4}, \delta M_{np}) U_{e 4}^2} -V_{ud,\rm n } \right]^2}{(\delta V_{ud,n})^2}\,. 
\end{align}
We have two different $\chi^2_{n}$ functions for the $V_{ud,n}= 0.97413(43)$ at the bottle UCN experiment \eqref{eq:bottle} and  
$V_{ud,n}= 0.96866(131)$ in the beam \eqref{eq:inbeam}.
We do not combine them due to their disagreement. 

Next, we construct $\chi^2$ function for the superallowed $0^+\to 0^+$ nuclear $\beta$ decay,
where the ten most accurate nuclei data are used.
Each superallowed transition is characterized by its own $f t$ value, where 
$f$ is a dimensionless constant that comes from an integral over phase space
and
$t$ is the half-lifetime of the $0^+\to 0^+$ transition~\cite{Hardy:2004dm,Hardy:2004id,Hardy:2008gy}.
By subtracting the nuclear structure-dependent corrections from each $f t$ value, one can obtain each corrected $\mathcal{F} t$ value, which is a nucleus-independent quantity according to the conserved vector current hypothesis, 
 and sensitive to $(G_F^{\rm obs}V_{ud}^{\rm obs})^{-2}$.
In the fit of ten superallowed nuclear decays, 
one needs a common nuisance parameter $\alpha$ to account for the total systematic uncertainty. 
With this consideration, $\chi^2$ functions from the superallowed transitions are given by
\begin{align}
&\chi^2_{0^+} (m_{\nu 4}, U_{e 4}^2, V_{ud}, \alpha)
\nonumber\\
&= \sum_i \left({\cal F}t_i -\frac{K (1+\Delta_R^V)^{-1}(1+\alpha)}{2 G_F^2 V_{ud}^2\left[ 1-\epsilon(m_{\nu 4}, \delta M_i) U_{e 4}^2\right]}  \right)^2/(\delta {\cal F}t_i)^2\,,
\\
&\chi^2_{\rm sys} (\alpha)=\frac{\alpha^2}{\sigma_{\Delta_R^V}^2+\sigma_{\delta'_R}^2+\sigma_{\delta_{\rm NS}}^2}\,,
\end{align}
where $G_F= 1.1663788(6) \times 10^{-5}\,\text{GeV}^{-2}$,
$K=8120.27648(26) \times 10^{-10}\,\text{GeV}^{-4}\text{s}$, and 
the index $i$ runs $^{10}$C,
$^{14}$O, 
$^{26m}$Al,
$^{34}$Cl,
$^{34}$Ar,
$^{38m}$K,
$^{42}$Sc,
$^{46}$V,
$^{50}$Mn, and 
$^{54}$Co, shown in Fig.~\ref{fig:phasespace}.
We use the transition-independent radiative correction $\Delta_R^V =2.467(27)\times 10^{-2}$ from Eq.~(A.8) of Ref.~\cite{Cirigliano:2022yyo},
giving $\sigma_{\Delta_R^V}=\delta \Delta_R^V/(1+ \Delta_R^V)=2.63\times 10^{-4}$. 
$\mathcal{F} t_i$ values are obtained from Table XVI of Ref.~\cite{Hardy:2020qwl}.
We extract $\sigma_{\delta'_R}=0.36\text{\,s}/3072 \text{\,s}=1.17\times 10^{-4}$ 
and $\sigma_{\delta_{\rm NS}}=1.73\text{\,s}/3072\text{\,s}=5.63\times 10^{-4}$ from Eq.~(22) of Ref.~\cite{Hardy:2020qwl}.

For a crosscheck of our statistical scheme of the superallowed nuclear decays,
when a sterile neutrino decoupling limit ($U_{e 4} \to 0$ or $m_{\nu 4}\to 0$) is examined with
$\Delta_R^V =2.454(19)\times 10^{-2}$ 
given in Ref.~\cite{Hardy:2020qwl}, 
we obtained $V_{ud,0^+}=0.97369(32)$,
which well agrees with  $V_{ud,0^+}= 0.97373(31)$
of Ref.~\cite{Hardy:2020qwl}. 

Finally, we combine all the $\chi^2$ functions
using only one neutron $\chi^2$ at a time, 
\begin{align}
\chi^2(m_{\nu 4}, U_{e 4}^2,  V_{ud}, \alpha) =
\chi_{us}^2 
+\chi_{us/ud}^2 
+\chi_{n}^2 
+  \chi_{0^+}^2 +\chi^2_{\rm sys}\,.
\end{align}
We find the global minimum $\chi^2_{\rm min}$ and search for the favored parameter space of the sterile neutrino by evaluating
\beq
\Delta \chi^2 \equiv \chi^2 (m_{\nu 4}, U_{e 4}^2,   V_{ud}, \alpha) -\chi^2_{\rm min}\,,
\eeq
where $V_{ud}$, as well as $\alpha$, is treated as the nuisance parameters and thus minimized for a given set of $(m_{\nu 4}, U_{e 4}^2)$.

The results are shown in Fig.~\ref{fig:neutrinofit}. 
As expected, the favored mixing angle squared is $U_{e 4}^2\approx 10^{-3}$,
and there is a plateau once the sterile neutrino mass $m_{\nu 4}$ is heavier than $\delta M-m_e\approx 3.2\,$MeV, 
which is the maximum neutrino energy at the superallowed nuclear decay  ${}^{26m}\text{Al}\to {}^{26}\text{Mg}$, see also Fig.~\ref{fig:phasespace}. 

We show two panels depending on the neutron lifetime measurements. In the left panel,  
we use the bottle measurement to extract $V_{ud}$, which is consistent with the one from the superallowed nuclear decays. The original tension of the CAA is at 2.8$\sigma$, and the pull in the presence of sterile neutrino is 2.1$\sigma$ at the best-fit point.
On the other hand, since the beam measurement prefers smaller $V_{ud}$ than the superallowed nuclear decays, 
a significance of the CAA is enhanced to be 3.8$\sigma$, and the sterile neutrino can relax it by  3.2$\sigma$ at the best-fit point.

\section{Constraints of MeV sterile neutrino}
\label{sec:constraint}

In the previous section, 
we identified the favored ranges of the mass and mixing of the sterile neutrino for the given anomaly.
However, these parameter regions are subject to constraints from the laboratory to the cosmology. 
In this section,
we list all the relevant bounds and show consistent resolutions if exist. 

\subsection{Direct bounds}
There have been kink searches in the Kurie plot, the emitted electron energy spectrum measurements in the nuclear $\beta$ decays, to test the mass of sterile neutrino \cite{Schreckenbach:1983cg,Deutsch:1990ut,Derbin:2018dbu,Bolton:2019pcu}. 
 One relevant measurement is the kink search in ${}^{20}{\rm F}$ decay which covers $m_{\nu 4}\lesssim 3$\,MeV \cite{Deutsch:1990ut}. 
The blue-shaded regions in Fig.~\ref{fig:neutrinofit} are excluded by the 
kink searches. 
This is a robust bound on the sterile neutrino.

Furthermore, recently the BeEST experiment has set a constraint on the sterile neutrino 
through the EC-$\beta$ decay using the superconducting quantum sensor \cite{Friedrich:2020nze}.
This bound is shown by the purple-shaded regions in Fig.~\ref{fig:neutrinofit}, which 
significantly improved the bound on the sterile neutrino mixing $U_{e4}$ with  $m_{\nu 4}=$0.1\,MeV\,--\,0.8\,MeV. 
This bound is also robust in our scenario. 

\subsection{\texorpdfstring{\boldmath{$0\nu\beta\beta$}}{0nu2beta} bound}

Searches for the neutrinoless double beta decay ($0\nu\beta\beta$) are sensitive to Majorana neutrino mass and mixing. See the recent review \cite{Bolton:2019pcu,Dekens:2023iyc}.
If a  $3+1$ sterile neutrino scenario (three active neutrinos plus one Majorana sterile neutrino) is considered, the $0\nu\beta\beta$ rate is proportional to $|U_{e 4}^2 m_{\nu 4}|^2$.
In this simplest case, the current measurements \cite{KamLAND-Zen:2016pfg,GERDA:2018pmc} set the upper limit on $U_{e4}^{2} < 10^{-7} $ at $m_{\nu 4}=\mathcal{O}(1)\,$MeV \cite{Bolton:2019pcu}, which excludes the whole parameter region of Fig.~\ref{fig:neutrinofit}.

However, it is known that the $0\nu\beta\beta$ bound can be suppressed by a generalized $B-L$ symmetry \cite{Chrzaszcz:2019inj}. In particular,
the bound is totally suppressed 
within the inverse seesaw model \cite{Mohapatra:1986bd,Mohapatra:1986aw,Nandi:1985uh} in this category. 
In this scenario, the left-handed singlet fermion $S$ as well as the right-handed neutrino $N$ are introduced to the SM Lagrangian
\cite{BhupalDev:2012jvh},
\beq
\!\!\!\! -\mathcal{L} &= 
 \frac{1}{2} \overline{N}^c \mu_N N
+ y \overline{L} \widetilde{H} N 
+ \frac{1}{2} \overline{S}^c \mu_S S
+ \overline{S} M_N N  + \textrm{h.c.}\,,
\eeq
where $\mu_N$ and $\mu_S$ are Majorana masses, which are small lepton number violations and are technically natural.
After taking the mass eigenbasis, the active neutrino mass is 
\beq 
m_{\nu'_\ell} \simeq \frac{M_D^2}{M_N^2} \mu_S\simeq U_{e 4}^2 \mu_S\,,
\eeq 
where the mixing is $U_{e4}^2 \simeq (M_D/M_N)^2$, while the sterile neutrino becomes a pseudo-Dirac fermion with mass as  
\beq
m_{\nu 4} \simeq M_N \pm \frac{\mu_S}{2}\,.
\eeq
Here, $M_N \gg M_D=v y ~(v\simeq 174\,$GeV) is assumed.
Note that $\mu_N$ is an irrelevant parameter at the leading order.

The smallness of the active neutrino masses requires $\mu_S$ to be suppressed in the inverse seesaw model, namely $\mu_s \approx \mathcal{O}(1)\,$eV in the parameter of our interest. Consequently, the $0\nu\beta\beta$ rate, which has to pick up the Majorana mass $\mu_s$, is significantly smaller than in the 3+1 scenario and  contained well below the observed limit for $m_{\nu 4}\lesssim \mathcal{O}(100)\,$MeV~\cite{Bolton:2019pcu}.

\subsection{Meson decay bounds}

  The sterile neutrino bounds from meson leptonic decays have been discussed in Ref.~\cite{Shrock:1980ct}. 
  Within the SM, $\pi^+ \to e^+ \nu$ decay is helicity-suppressed by the electron mass ($m_e^2/m_{\pi^+}^2$).
  On the other hand, when the sterile neutrino is heavier than the electron, 
 the branching ratio is significantly modified. 
 Strong constraints come from 
the $e$-$\mu$ universality measurements in the two-body leptonic decays of $\pi^+$ and $K^+$ \cite{Shrock:1980ct,Bryman:1983cja,Britton:1992pg,Britton:1992xv,Abada:2012mc,Abada:2013aba},
\begin{align}
  R_{e/\mu}^{M} =   \frac{\text{BR}(M^+ \to e^+ \nu (\gamma))}{\text{BR}(M^+ \to \mu^+ \nu (\gamma))}\,, \quad \text{for~}M=\pi,\,K.
\end{align}
In Fig.~\ref{fig:neutrinofit}, the areas above the green dashed lines are excluded by the latest analysis of $R_{e/\mu}^{\pi}$ at the PiENu experiment \cite{PiENu:2015seu,Bryman:2019ssi,Bryman:2019bjg} (the bound from $R_{e/\mu}^{K}$ is weaker \cite{NA62:2012lny}).  
 In the minimal scenario of the sterile neutrino with the mass and mixing to the weak interaction, 
 this bound excludes the significant part of the favored parameter space for the CAA.\footnote{%
 The introduction of the muon-neutrino mixing ($U_{\mu 4}$) does not ameliorate the situation. 
 It is because the effect to  $R_{e/\mu}^{M}$ from $U_{e4}^2$ is chirality enhanced  by a factor of $m_{\nu 4}^2/m_e^2$. In contrast, the effect from $U_{\mu 4}^2$ is diminished, being suppressed by $m_{\nu 4}^2/m_M^2$ for $M=\pi,\,K$~\cite{Bryman:2019bjg}.}

 However, if there are some higher dimensional operators that contain the sterile neutrino and induce $\pi^+\to e^+ \nu_{4}$, they can reduce $\pi^+\to e^+ \nu_{4}$ while keeping the nuclear and neutron $\beta$ decay unaffected.
 This is because the SM-like amplitude of the pion decay is actually suppressed by the lepton mass, 
 so its electron mode is sensitive to new physics contributions.
Even with new physics of $\mathcal{O}(10)$\,TeV scale, the $R_{e/\mu}^{\pi}$ bound can be easily compensated by small modification due to the higher dimensional operators.

 For example, $R_{e/\mu}^{M}$  is sensitive to a dimension-six scalar operator 
 $(\overline{u}_R V_{ui} d_L^i)(\overline{e}_L N)$,
which can destructively interfere with the mixing contribution of Eq.~\eqref{eq:mixing}, and we introduce 
 \begin{align}
\mathcal{L}_{\rm eff}= \frac{1}{\Lambda^2}\left[\overline{u}_{R} (V_{ud}d_L + V_{us}s_L)\right](\overline{e}_L  N) + \textrm{h.c.}\,.
\label{eq:dim6}
 \end{align}
In Fig.~\ref{fig:neutrinofit}, 
we show that modified allowed regions of  $R_{e/\mu}^{M}$ for the cases of $\Lambda = $20\,TeV and 30\,TeV by the green dash-dotted  and dotted lines, respectively.
It is found that the $R_{e/\mu}^{M}$ bound is significantly sensitive to the operator of Eq.~\eqref{eq:dim6}, and the CAA favors the dimension-six operator with $\Lambda=$20\,--\,30\,TeV.

\subsection{Long-lived sterile neutrino and cosmology}

When the sterile neutrino lifetime is determined by the weak decay, 
the reactor \cite{Derbin:1993wy,Hagner:1995bn} and Borexino experiments \cite{Borexino:2013bot} use  $\nu_{4}  \to e^+ e^- \nu'_e$ decay mode to probe  the range of 
1\,MeV $< m_{\nu 4}<$ 14\,MeV \cite{Bolton:2019pcu}.
These bounds potentially exclude the favored regions of Fig.~\ref{fig:neutrinofit}.
However, the bounds are not applied if the lifetime is shorter by a factor of $\mathcal{O}(10-100)$ due to the additional decay modes.

The decay of sterile neutrino to three active neutrinos becomes significant in the presence of a real scalar mediator $\phi$  described by 
\beq
-\mathcal{L}=  \frac{m_\phi^2}{2} \phi^2
+\left(\frac{\lambda}{2} \phi \overline{S}^c  S 
+ \text{h.c.}\right)\,.
\eeq
If the effective interaction scale  satisfies  $\lambda^2/m_\phi^2 \gtrsim 10 \,U_{e4}^{-2} G_F  \sim 0.1/{\rm GeV}^2$, the decay of the sterile neutrino is short enough. We find, when $m_\phi\simeq 0.5$\,GeV\,--\,30\,GeV,  this condition is compatible with the bounds, such as the meson decays, studied in Ref.~\cite{deGouvea:2019qaz} since the mediator interactions with the active-neutrino are suppressed by the mixing angle as $\frac{1}{2}\lambda U_{e 4}^2 \phi \bar \nu_e^c \nu_e, \lambda U_{e 4} \phi \overline{S}^c \nu_e $.

The cosmological observations typically constrain the MeV scale sterile neutrino. However, with the required mediator-interaction, the sterile neutrino decoupling from the SM thermal bath occurs together with the active neutrinos decoupling at $T\sim 2$\,MeV. As a result, the remaining bound is from the effective number of neutrinos, $N_{\rm eff}$. As the sterile neutrinos are not completely non-relativistic at the decoupling temperature, they would increase the effective number of neutrinos by $\Delta N_{\rm eff}\simeq 1.0\text{\,--\,}0.25$ for $m_{\nu 4}=5$\,--\,10\,MeV. 
Although  $\Delta N_{\rm eff}\gtrsim 0.3$ is constrained by the CMB observations \cite{Planck:2018vyg}, additional well-motivated particles, such as heavy axions with a lifetime of about 0.1\,--\,1 sec, can consistently compensate  $\Delta N_{\rm eff}$~\cite{Cadamuro:2010cz, Dunsky:2022uoq}. 

In the absence of additional long-lived particles affecting $N_{\rm eff}$, a different set of mediator interactions, 
\beq
-\mathcal{L}_\phi = \frac{m_\phi^2}{2} \phi^2 + \frac{\phi}{4\Lambda_\gamma} F_{\mu\nu} F^{\mu \nu}+\left( \lambda' \phi \overline{S}_L N  
+ \text{h.c.}\right)\,,
\eeq
can lead to another consistent scenario. In this case, the sterile neutrino decay is predominantly $\nu_{4} \to \phi^{(\ast)} \nu'_e \to  \gamma \gamma \nu'_e$ which dismisses the reactor and Borexino bounds. 
The thermal history is modified such that the sterile neutrino interaction with the photons determines the decoupling from the thermal bath. The temperature when the entire neutrino sector decouples can be lower to $1(0.75)$\,MeV, which is about the temperature of neutron decoupling. The sterile neutrino at this temperature is non-relativistic enough such that  $\Delta N_{\rm eff}\simeq 0.3\text{\,--\,}9\times 10^{-3} (0.1\text{\,--\,}6\times 10^{-4})$  for $m_{\nu 4}=5$\,--\,10\,MeV is allowed by the CMB.  
This scenario requires $\lambda'/ (m_\phi^2)\Lambda\gtrsim 0.03/\text{GeV}^3$. 
Considering the bounds involving  neutrino \cite{deGouvea:2019qaz} and photon \cite{Dolan:2017osp, Gori:2020xvq}, we find that  the mediator is allowed when $m_\phi\simeq 0.5$\,GeV\,--\,4\,GeV and $\lambda^\prime \gtrsim 1$. 

\subsection{PMNS unitarity test}
The Pontecorvo-Maki-Nakagawa-Sakata (PMNS) matrix~\cite{Pontecorvo:1957qd,Maki:1962mu} is a unitary matrix relating the mass eigenstates of neutrinos to the charged-lepton flavor eigenstates. Although the PMNS matrix is usually defined within three active neutrinos, the presence of sterile neutrinos extends the matrix and makes the $3\times 3$ sub-matrix non-unitary.
Hence, the unitarity test on the sub-matrix with consideration of neutrino oscillation data constraints the mixing elements due to the sterile neutrinos~\cite{Parke:2015goa,Fong:2016yyh,Ellis:2020hus,Hu:2020oba, Goldhagen:2021kxe}. The recent global analysis  suggests  $|U_{e4}|^2\lesssim 0.03$ at 2$\sigma$~\cite{Goldhagen:2021kxe},
which is sufficiently weaker than the size of $|U_{e4}|^2$ favored by the CAA within our scenario.

\subsection{Summary of the constraints}
Given the potential constraints for the parameter space favored by the CAA, the inverse seesaw models with a mediator $\phi$ and a higher dimensional operator emerge as viable scenarios.  
Even in the presence of the additional decay modes of the sterile neutrino via the mediator, it remains effectively stable in most laboratory experiments, in particular, measurements of neutron and nuclear $\beta$ decays.

\section{Conclusions}
\label{sec:conclusion}

Currently, it is reported that the first-row CKM unitarity test is violated at the $2.8\sigma$ level, referred to as the Cabibbo angle anomaly (CAA). 
This violation is worse when the neutron lifetime data of the in-beam experiment is used in the global fit.
In this article, we point out that the MeV sterile neutrino that mixes the electron-type neutrino with $U_{e 4}^2\approx 10^{-3}$ can decrease the value of extracted $|V_{ud}^{\rm obs}|$ from the superallowed nuclear $\beta$ decays and the neutron decay relative to its true value, without modifying other observables relevant to the CKM determinations.
As a result, it is found that the MeV sterile neutrino can alleviate the CAA.
The parameter space of the sterile neutrino favored as the solution of the CAA is typically subject to various constraints. 
Although the nuclear $\beta$ decay bounds are robust, we show that the sterile neutrino in the inverse seesaw models with the dimension-six operators can evade the laboratory constraints,
and the neutrino mediator is favored by the cosmology bounds.

While we highlighted the sterile neutrino resolving the anomaly of the unitarity test, this type of scenario where new physics primarily in the neutrino sector gives apparent deviations in the precision measurements on the quark sector could be interesting in a broader context.

\section*{Note Added}
After completing this work, a new lattice calculation~\cite{Ma:2023kfr} reported $\Delta_R^V=0.02439(19)$. 
If we adopt this result for the global fit of the superallowed nuclear $\beta$ decays,
significance of the tension is reduced by about 0.5$\sigma$, and correspondingly, the improvement of CAA (the pull with respect to the SM) due to the sterile neutrino is reduced by about 0.5$\sigma$.  
\acknowledgments

We would like to thank  Daniel Ega\~{n}a-Ugrinovic, Matheus Hostert, Takemichi Okui, Robert Shrock, Ningqiang Song, Mark-Christoph Spieker,  Vandana Tripathi, and Nodoka Yamanaka for useful discussions. 
T.\,K.\ was supported by the Japan Society for the Promotion of Science (JSPS) Grant-in-Aid for Scientific Research  (Grant No.\,21K03572) and the JSPS Core-to-Core Program (Grant No.\,JPJSCCA20200002). 
K.\,T.\ is supported by in part the
US Department of Energy grant DE-SC0010102 and JSPS 
Grant-in-Aid for Scientific Research (Grant No.\,21H01086).

\bibliographystyle{utphys28mod}
\bibliography{ref}

\providecommand{\href}[2]{#2}\begingroup\raggedright\begin{thebibliography}{10}

\bibitem{HFLAV:2022pwe}
{\bfseries Heavy Flavor Averaging Group, HFLAV} Collaboration, ``{Averages of
  $b$-hadron, $c$-hadron, and $\tau$-lepton properties as of 2021},''
  \href{https://dx.doi.org/10.1103/PhysRevD.107.052008}{Phys.\  Rev.\  D
  {\bfseries 107} (2023) 052008} {\ttfamily
  [\href{https://arxiv.org/abs/2206.07501}{arXiv:2206.07501}]}.

\bibitem{Cabibbo:1963yz}
N.~Cabibbo, ``{Unitary Symmetry and Leptonic Decays},''
  \href{https://dx.doi.org/10.1103/PhysRevLett.10.531}{Phys.\  Rev.\  Lett.\
  {\bfseries 10} (1963) 531--533}.

\bibitem{Kobayashi:1973fv}
M.~Kobayashi and T.~Maskawa, ``{CP Violation in the Renormalizable Theory of
  Weak Interaction},'' \href{https://dx.doi.org/10.1143/PTP.49.652}{Prog.\
  Theor.\  Phys.\  {\bfseries 49} (1973) 652--657}.

\bibitem{Seng:2018yzq}
C.-Y.~Seng, M.~Gorchtein, H.~H.~Patel, and M.~J.~Ramsey-Musolf, ``{Reduced
  Hadronic Uncertainty in the Determination of $V_{ud}$},''
  \href{https://dx.doi.org/10.1103/PhysRevLett.121.241804}{Phys.\  Rev.\
  Lett.\  {\bfseries 121} (2018) 241804} {\ttfamily
  [\href{https://arxiv.org/abs/1807.10197}{arXiv:1807.10197}]}.

\bibitem{Belfatto:2019swo}
B.~Belfatto, R.~Beradze, and Z.~Berezhiani, ``{The CKM unitarity problem: A
  trace of new physics at the TeV scale?}''
  \href{https://dx.doi.org/10.1140/epjc/s10052-020-7691-6}{Eur.\  Phys.\  J.\
  C {\bfseries 80} (2020) 149} {\ttfamily
  [\href{https://arxiv.org/abs/1906.02714}{arXiv:1906.02714}]}.

\bibitem{Grossman:2019bzp}
Y.~Grossman, E.~Passemar, and S.~Schacht, ``{On the Statistical Treatment of
  the Cabibbo Angle Anomaly},''
  \href{https://dx.doi.org/10.1007/JHEP07(2020)068}{JHEP {\bfseries 07} (2020)
  068} {\ttfamily [\href{https://arxiv.org/abs/1911.07821}{arXiv:1911.07821}]}.

\bibitem{Coutinho:2019aiy}
A.~M.~Coutinho, A.~Crivellin, and C.~A.~Manzari, ``{Global Fit to Modified
  Neutrino Couplings and the Cabibbo-Angle Anomaly},''
  \href{https://dx.doi.org/10.1103/PhysRevLett.125.071802}{Phys.\  Rev.\
  Lett.\  {\bfseries 125} (2020) 071802} {\ttfamily
  [\href{https://arxiv.org/abs/1912.08823}{arXiv:1912.08823}]}.

\bibitem{Cheung:2020vqm}
K.~Cheung, W.-Y.~Keung, C.-T.~Lu, and P.-Y.~Tseng, ``{Vector-like Quark
  Interpretation for the CKM Unitarity Violation, Excess in Higgs Signal
  Strength, and Bottom Quark Forward-Backward Asymmetry},''
  \href{https://dx.doi.org/10.1007/JHEP05(2020)117}{JHEP {\bfseries 05} (2020)
  117} {\ttfamily [\href{https://arxiv.org/abs/2001.02853}{arXiv:2001.02853}]}.

\bibitem{Seng:2020wjq}
C.-Y.~Seng, X.~Feng, M.~Gorchtein, and L.-C.~Jin, ``{Joint lattice
  QCD\textendash{}dispersion theory analysis confirms the quark-mixing top-row
  unitarity deficit},''
  \href{https://dx.doi.org/10.1103/PhysRevD.101.111301}{Phys.\  Rev.\  D
  {\bfseries 101} (2020) 111301} {\ttfamily
  [\href{https://arxiv.org/abs/2003.11264}{arXiv:2003.11264}]}.

\bibitem{Crivellin:2020ebi}
A.~Crivellin, F.~Kirk, C.~A.~Manzari, and M.~Montull, ``{Global Electroweak Fit
  and Vector-Like Leptons in Light of the Cabibbo Angle Anomaly},''
  \href{https://dx.doi.org/10.1007/JHEP12(2020)166}{JHEP {\bfseries 12} (2020)
  166} {\ttfamily [\href{https://arxiv.org/abs/2008.01113}{arXiv:2008.01113}]}.

\bibitem{Kirk:2020wdk}
M.~Kirk, ``{Cabibbo anomaly versus electroweak precision tests: An exploration
  of extensions of the Standard Model},''
  \href{https://dx.doi.org/10.1103/PhysRevD.103.035004}{Phys.\  Rev.\  D
  {\bfseries 103} (2021) 035004} {\ttfamily
  [\href{https://arxiv.org/abs/2008.03261}{arXiv:2008.03261}]}.

\bibitem{Crivellin:2021njn}
A.~Crivellin, M.~Hoferichter, and C.~A.~Manzari, ``{Fermi Constant from Muon
  Decay Versus Electroweak Fits and Cabibbo-Kobayashi-Maskawa Unitarity},''
  \href{https://dx.doi.org/10.1103/PhysRevLett.127.071801}{Phys.\  Rev.\
  Lett.\  {\bfseries 127} (2021) 071801} {\ttfamily
  [\href{https://arxiv.org/abs/2102.02825}{arXiv:2102.02825}]}.

\bibitem{Belfatto:2021jhf}
B.~Belfatto and Z.~Berezhiani, ``{Are the CKM anomalies induced by vector-like
  quarks? Limits from flavor changing and Standard Model precision tests},''
  \href{https://dx.doi.org/10.1007/JHEP10(2021)079}{JHEP {\bfseries 10} (2021)
  079} {\ttfamily [\href{https://arxiv.org/abs/2103.05549}{arXiv:2103.05549}]}.

\bibitem{Bryman:2021teu}
D.~Bryman, V.~Cirigliano, A.~Crivellin, and G.~Inguglia, ``{Testing Lepton
  Flavor Universality with Pion, Kaon, Tau, and Beta Decays},''
  \href{https://dx.doi.org/10.1146/annurev-nucl-110121-051223}{Ann.\  Rev.\
  Nucl.\  Part.\  Sci.\  {\bfseries 72} (2022) 69--91} {\ttfamily
  [\href{https://arxiv.org/abs/2111.05338}{arXiv:2111.05338}]}.

\bibitem{Cirigliano:2022yyo}
V.~Cirigliano, A.~Crivellin, M.~Hoferichter, and M.~Moulson, ``{Scrutinizing
  CKM unitarity with a new measurement of the $K_{\mu 3}/K_{\mu 2}$ branching
  fraction},'' \href{https://dx.doi.org/10.1016/j.physletb.2023.137748}{Phys.\
  Lett.\  B {\bfseries 838} (2023) 137748} {\ttfamily
  [\href{https://arxiv.org/abs/2208.11707}{arXiv:2208.11707}]}.

\bibitem{Crivellin:2022rhw}
A.~Crivellin, M.~Kirk, T.~Kitahara, and F.~Mescia, ``{Global fit of modified
  quark couplings to EW gauge bosons and vector-like quarks in light of the
  Cabibbo angle anomaly},''
  \href{https://dx.doi.org/10.1007/JHEP03(2023)234}{JHEP {\bfseries 03} (2023)
  234} {\ttfamily [\href{https://arxiv.org/abs/2212.06862}{arXiv:2212.06862}]}.

\bibitem{Belfatto:2023tbv}
B.~Belfatto and S.~Trifinopoulos, ``{The remarkable role of the vector-like
  quark doublet in the Cabibbo angle and $W$-mass anomalies}.'' {\ttfamily
  \href{https://arxiv.org/abs/2302.14097}{arXiv:2302.14097}}.

\bibitem{UCNt:2021pcg}
{\bfseries UCN$\tau$} Collaboration, ``{Improved neutron lifetime measurement
  with UCN$\tau$},''
  \href{https://dx.doi.org/10.1103/PhysRevLett.127.162501}{Phys.\  Rev.\
  Lett.\  {\bfseries 127} (2021) 162501} {\ttfamily
  [\href{https://arxiv.org/abs/2106.10375}{arXiv:2106.10375}]}.

\bibitem{Yue:2013qrc}
A.~T.~Yue, {\em et al.}, ``{Improved Determination of the Neutron Lifetime},''
  \href{https://dx.doi.org/10.1103/PhysRevLett.111.222501}{Phys.\  Rev.\
  Lett.\  {\bfseries 111} (2013) 222501} {\ttfamily
  [\href{https://arxiv.org/abs/1309.2623}{arXiv:1309.2623}]}.

\bibitem{Markisch:2018ndu}
B.~M\"arkisch {\em et~al.}, ``{Measurement of the Weak Axial-Vector Coupling
  Constant in the Decay of Free Neutrons Using a Pulsed Cold Neutron Beam},''
  \href{https://dx.doi.org/10.1103/PhysRevLett.122.242501}{Phys.\  Rev.\
  Lett.\  {\bfseries 122} (2019) 242501} {\ttfamily
  [\href{https://arxiv.org/abs/1812.04666}{arXiv:1812.04666}]}.

\bibitem{Charles:2004jd}
{\bfseries CKMfitter Group} Collaboration, ``{CP violation and the CKM matrix:
  Assessing the impact of the asymmetric $B$ factories},''
  \href{https://dx.doi.org/10.1140/epjc/s2005-02169-1}{Eur.\  Phys.\  J.\  C
  {\bfseries 41} (2005) 1--131} {\ttfamily
  [\href{https://arxiv.org/abs/hep-ph/0406184}{hep-ph/0406184}]}. {updated
  results and plots available at: \url{http://ckmfitter.in2p3.fr}}.

\bibitem{UTfit:2022hsi}
{\bfseries UTfit} Collaboration, ``{New UTfit Analysis of the Unitarity
  Triangle in the Cabibbo-Kobayashi-Maskawa scheme},''
  \href{https://dx.doi.org/10.1007/s12210-023-01137-5}{Rend.\  Lincei Sci.\
  Fis.\  Nat.\  {\bfseries 34} (2023) 37--57} {\ttfamily
  [\href{https://arxiv.org/abs/2212.03894}{arXiv:2212.03894}]}.

\bibitem{Crivellin:2020lzu}
A.~Crivellin and M.~Hoferichter, ``{\ensuremath{\beta} Decays as Sensitive
  Probes of Lepton Flavor Universality},''
  \href{https://dx.doi.org/10.1103/PhysRevLett.125.111801}{Phys.\  Rev.\
  Lett.\  {\bfseries 125} (2020) 111801} {\ttfamily
  [\href{https://arxiv.org/abs/2002.07184}{arXiv:2002.07184}]}.

\bibitem{Endo:2020tkb}
M.~Endo and S.~Mishima, ``{Muon $g-2$ and CKM unitarity in extra lepton
  models},'' \href{https://dx.doi.org/10.1007/JHEP08(2020)004}{JHEP {\bfseries
  08} (2020) 004} {\ttfamily
  [\href{https://arxiv.org/abs/2005.03933}{arXiv:2005.03933}]}.

\bibitem{Capdevila:2020rrl}
B.~Capdevila, A.~Crivellin, C.~A.~Manzari, and M.~Montull, ``{Explaining $b\to
  s\ell^+\ell^-$ and the Cabibbo angle anomaly with a vector triplet},''
  \href{https://dx.doi.org/10.1103/PhysRevD.103.015032}{Phys.\  Rev.\  D
  {\bfseries 103} (2021) 015032} {\ttfamily
  [\href{https://arxiv.org/abs/2005.13542}{arXiv:2005.13542}]}.

\bibitem{Li:2020wxi}
T.~Li, X.-D.~Ma, and M.~A.~Schmidt, ``{Constraints on the charged currents in
  general neutrino interactions with sterile neutrinos},''
  \href{https://dx.doi.org/10.1007/JHEP10(2020)115}{JHEP {\bfseries 10} (2020)
  115} {\ttfamily [\href{https://arxiv.org/abs/2007.15408}{arXiv:2007.15408}]}.

\bibitem{Crivellin:2020oup}
A.~Crivellin, C.~A.~Manzari, M.~Alguero, and J.~Matias, ``{Combined Explanation
  of the Z\textrightarrow{}bb\textasciimacron{} Forward-Backward Asymmetry, the
  Cabibbo Angle Anomaly, and
  \ensuremath{\tau}\textrightarrow{}\ensuremath{\mu}\ensuremath{\nu}\ensuremath{\nu}
  and b\textrightarrow{}s\ensuremath{\ell}+\ensuremath{\ell}- Data},''
  \href{https://dx.doi.org/10.1103/PhysRevLett.127.011801}{Phys.\  Rev.\
  Lett.\  {\bfseries 127} (2021) 011801} {\ttfamily
  [\href{https://arxiv.org/abs/2010.14504}{arXiv:2010.14504}]}.

\bibitem{Crivellin:2020klg}
A.~Crivellin, F.~Kirk, C.~A.~Manzari, and L.~Panizzi, ``{Searching for lepton
  flavor universality violation and collider signals from a singly charged
  scalar singlet},''
  \href{https://dx.doi.org/10.1103/PhysRevD.103.073002}{Phys.\  Rev.\  D
  {\bfseries 103} (2021) 073002} {\ttfamily
  [\href{https://arxiv.org/abs/2012.09845}{arXiv:2012.09845}]}.

\bibitem{Felkl:2021qdn}
T.~Felkl, J.~Herrero-Garcia, and M.~A.~Schmidt, ``{The Singly-Charged Scalar
  Singlet as the Origin of Neutrino Masses},''
  \href{https://dx.doi.org/10.1007/JHEP05(2021)122}{JHEP {\bfseries 05} (2021)
  122} {\ttfamily [\href{https://arxiv.org/abs/2102.09898}{arXiv:2102.09898}]}.
  [Erratum: JHEP 05, 073 (2022)].

\bibitem{Branco:2021vhs}
G.~C.~Branco, J.~T.~Penedo, P.~M.~F.~Pereira, M.~N.~Rebelo, and
  J.~I.~Silva-Marcos, ``{Addressing the CKM unitarity problem with a
  vector-like up quark},''
  \href{https://dx.doi.org/10.1007/JHEP07(2021)099}{JHEP {\bfseries 07} (2021)
  099} {\ttfamily [\href{https://arxiv.org/abs/2103.13409}{arXiv:2103.13409}]}.

\bibitem{Marzocca:2021azj}
D.~Marzocca and S.~Trifinopoulos, ``{Minimal Explanation of Flavor Anomalies:
  B-Meson Decays, Muon Magnetic Moment, and the Cabibbo Angle},''
  \href{https://dx.doi.org/10.1103/PhysRevLett.127.061803}{Phys.\  Rev.\
  Lett.\  {\bfseries 127} (2021) 061803} {\ttfamily
  [\href{https://arxiv.org/abs/2104.05730}{arXiv:2104.05730}]}.

\bibitem{Buras:2021btx}
A.~J.~Buras, A.~Crivellin, F.~Kirk, C.~A.~Manzari, and M.~Montull, ``{Global
  analysis of leptophilic Z' bosons},''
  \href{https://dx.doi.org/10.1007/JHEP06(2021)068}{JHEP {\bfseries 06} (2021)
  068} {\ttfamily [\href{https://arxiv.org/abs/2104.07680}{arXiv:2104.07680}]}.

\bibitem{Crivellin:2021bkd}
A.~Crivellin, M.~Hoferichter, M.~Kirk, C.~A.~Manzari, and L.~Schnell,
  ``{First-generation new physics in simplified models: from low-energy parity
  violation to the LHC},''
  \href{https://dx.doi.org/10.1007/JHEP10(2021)221}{JHEP {\bfseries 10} (2021)
  221} {\ttfamily [\href{https://arxiv.org/abs/2107.13569}{arXiv:2107.13569}]}.

\bibitem{Cirigliano:2021yto}
V.~Cirigliano, D.~D\'\i{}az-Calder\'on, A.~Falkowski, M.~Gonz\'alez-Alonso, and
  A.~Rodr\'\i{}guez-S\'anchez, ``{Semileptonic tau decays beyond the Standard
  Model},'' \href{https://dx.doi.org/10.1007/JHEP04(2022)152}{JHEP {\bfseries
  04} (2022) 152} {\ttfamily
  [\href{https://arxiv.org/abs/2112.02087}{arXiv:2112.02087}]}.

\bibitem{Blennow:2022yfm}
M.~Blennow, P.~Coloma, E.~Fern\'andez-Mart\'\i{}nez, and M.~Gonz\'alez-L\'opez,
  ``{Right-handed neutrinos and the CDF II anomaly},''
  \href{https://dx.doi.org/10.1103/PhysRevD.106.073005}{Phys.\  Rev.\  D
  {\bfseries 106} (2022) 073005} {\ttfamily
  [\href{https://arxiv.org/abs/2204.04559}{arXiv:2204.04559}]}.

\bibitem{Bryman:2019bjg}
D.~A.~Bryman and R.~Shrock, ``{Constraints on Sterile Neutrinos in the MeV to
  GeV Mass Range},''
  \href{https://dx.doi.org/10.1103/PhysRevD.100.073011}{Phys.\  Rev.\  D
  {\bfseries 100} (2019) 073011} {\ttfamily
  [\href{https://arxiv.org/abs/1909.11198}{arXiv:1909.11198}]}.

\bibitem{Kitahara:2023kmh}
T.~Kitahara and K.~Tobioka, ``{Sterile neutrinos in light of the Cabibbo-angle
  anomaly},'' \href{https://dx.doi.org/10.1088/1742-6596/2446/1/012009}{J.\
  Phys.\  Conf.\  Ser.\  {\bfseries 2446} (2023) 012009}.

\bibitem{CDF:2022hxs}
{\bfseries CDF} Collaboration, ``{High-precision measurement of the W boson
  mass with the CDF II detector},''
  \href{https://dx.doi.org/10.1126/science.abk1781}{Science {\bfseries 376}
  (2022) 170--176}.

\bibitem{Hardy:2020qwl}
J.~C.~Hardy and I.~S.~Towner, ``{Superallowed $0^+ \to 0^+$ nuclear $\beta$
  decays: 2020 critical survey, with implications for V$_{ud}$ and CKM
  unitarity},'' \href{https://dx.doi.org/10.1103/PhysRevC.102.045501}{Phys.\
  Rev.\  C {\bfseries 102} (2020) 045501}.

\bibitem{Brindhaban:1994zz}
S.~A.~Brindhaban and P.~H.~Barker, ``{The Q value for the Alm26 superallowed
  beta decay},'' \href{https://dx.doi.org/10.1103/PhysRevC.49.2401}{Phys.\
  Rev.\  C {\bfseries 49} (1994) 2401--2406}.

\bibitem{Eronen:2006if}
T.~Eronen {\em et~al.}, ``{Q-values of the Superallowed beta-Emitters Al-26-m,
  Sc-42 and V-46 and their impact on V(ud) and the Unitarity of the CKM
  Matrix},'' \href{https://dx.doi.org/10.1103/PhysRevLett.97.232501}{Phys.\
  Rev.\  Lett.\  {\bfseries 97} (2006) 232501} {\ttfamily
  [\href{https://arxiv.org/abs/nucl-ex/0606035}{nucl-ex/0606035}]}.

\bibitem{NIST_ASD}
A.~Kramida, {Yu. ~Ralchenko}, J.~Reader, and {NIST ASD Team}. {``NIST Atomic
  Spectra Database (ver. 5.10),'' Available:
  {\url{https://www.nist.gov/pml/atomic-spectra-database}}. National Institute
  of Standards and Technology, Gaithersburg, MD.}, 2022.
\newblock \href{https://dx.doi.org/10.18434/T4W30F}{{\ttfamily
  doi:10.18434/T4W30F}}.

\bibitem{Kleesiek:2018mel}
M.~Kleesiek {\em et~al.}, ``{$\beta$-Decay Spectrum, Response Function and
  Statistical Model for Neutrino Mass Measurements with the KATRIN
  Experiment},'' \href{https://dx.doi.org/10.1140/epjc/s10052-019-6686-7}{Eur.\
   Phys.\  J.\  C {\bfseries 79} (2019) 204} {\ttfamily
  [\href{https://arxiv.org/abs/1806.00369}{arXiv:1806.00369}]}.

\bibitem{Czarnecki:2018okw}
A.~Czarnecki, W.~J.~Marciano, and A.~Sirlin, ``{Neutron Lifetime and Axial
  Coupling Connection},''
  \href{https://dx.doi.org/10.1103/PhysRevLett.120.202002}{Phys.\  Rev.\
  Lett.\  {\bfseries 120} (2018) 202002} {\ttfamily
  [\href{https://arxiv.org/abs/1802.01804}{arXiv:1802.01804}]}.

\bibitem{Schreckenbach:1983cg}
K.~Schreckenbach, G.~Colvin, and F.~Von~Feilitzsch, ``{Search for mixing of
  heavy neutrinos in the $\beta^+$ and $\beta^-$ spectra of the $^{64}$Cu
  decay},'' \href{https://dx.doi.org/10.1016/0370-2693(83)90858-4}{Phys.\
  Lett.\  B {\bfseries 129} (1983) 265--268}.

\bibitem{Deutsch:1990ut}
J.~Deutsch, M.~Lebrun, and R.~Prieels, ``{Searches for admixture of massive
  neutrinos into the electron flavor},''
  \href{https://dx.doi.org/10.1016/0375-9474(90)90541-S}{Nucl.\  Phys.\  A
  {\bfseries 518} (1990) 149--155}.

\bibitem{Derbin:2018dbu}
A.~V.~Derbin, {\em et al.}, ``{Search for a Neutrino with a Mass of 0.01--1.0
  MeV in Beta Decays of $^{144}$Ce--$^{144}$Pr Nuclei},''
  \href{https://dx.doi.org/10.1134/S0021364018200067}{JETP Lett.\  {\bfseries
  108} (2018) 499--503}.

\bibitem{Bolton:2019pcu}
P.~D.~Bolton, F.~F.~Deppisch, and P.~S.~Bhupal~Dev, ``{Neutrinoless double beta
  decay versus other probes of heavy sterile neutrinos},''
  \href{https://dx.doi.org/10.1007/JHEP03(2020)170}{JHEP {\bfseries 03} (2020)
  170} {\ttfamily [\href{https://arxiv.org/abs/1912.03058}{arXiv:1912.03058}]}.

\bibitem{Friedrich:2020nze}
S.~Friedrich {\em et~al.}, ``{Limits on the Existence of sub-MeV Sterile
  Neutrinos from the Decay of $^7$Be in Superconducting Quantum Sensors},''
  \href{https://dx.doi.org/10.1103/PhysRevLett.126.021803}{Phys.\  Rev.\
  Lett.\  {\bfseries 126} (2021) 021803} {\ttfamily
  [\href{https://arxiv.org/abs/2010.09603}{arXiv:2010.09603}]}.

\bibitem{PiENu:2015seu}
{\bfseries PiENu} Collaboration, ``{Improved Measurement of the $\pi \to
  \textrm{e} \nu$ Branching Ratio},''
  \href{https://dx.doi.org/10.1103/PhysRevLett.115.071801}{Phys.\  Rev.\
  Lett.\  {\bfseries 115} (2015) 071801} {\ttfamily
  [\href{https://arxiv.org/abs/1506.05845}{arXiv:1506.05845}]}.

\bibitem{Bryman:2019ssi}
D.~A.~Bryman and R.~Shrock, ``{Improved Constraints on Sterile Neutrinos in the
  MeV to GeV Mass Range},''
  \href{https://dx.doi.org/10.1103/PhysRevD.100.053006}{Phys.\  Rev.\  D
  {\bfseries 100} (2019) 053006} {\ttfamily
  [\href{https://arxiv.org/abs/1904.06787}{arXiv:1904.06787}]}.

\bibitem{Hardy:2004dm}
J.~C.~Hardy and I.~S.~Towner, ``{New limit on fundamental weak-interaction
  parameters from superallowed beta decay},''
  \href{https://dx.doi.org/10.1103/PhysRevLett.94.092502}{Phys.\  Rev.\  Lett.\
   {\bfseries 94} (2005) 092502} {\ttfamily
  [\href{https://arxiv.org/abs/nucl-th/0412050}{nucl-th/0412050}]}.

\bibitem{Hardy:2004id}
J.~C.~Hardy and I.~S.~Towner, ``{Superallowed $0^+ \to 0^+$ nuclear beta
  decays: A Critical survey with tests of CVC and the standard model},''
  \href{https://dx.doi.org/10.1103/PhysRevC.71.055501}{Phys.\  Rev.\  C
  {\bfseries 71} (2005) 055501} {\ttfamily
  [\href{https://arxiv.org/abs/nucl-th/0412056}{nucl-th/0412056}]}.

\bibitem{Hardy:2008gy}
J.~C.~Hardy and I.~S.~Towner, ``{Superallowed $0^+ \to 0^+$ nuclear beta
  decays: A New survey with precision tests of the conserved vector current
  hypothesis and the standard model},''
  \href{https://dx.doi.org/10.1103/PhysRevC.79.055502}{Phys.\  Rev.\  C
  {\bfseries 79} (2009) 055502} {\ttfamily
  [\href{https://arxiv.org/abs/0812.1202}{arXiv:0812.1202}]}.

\bibitem{Dekens:2023iyc}
W.~Dekens, {\em et al.}, ``{Neutrinoless double-beta decay in the
  neutrino-extended Standard Model}.'' {\ttfamily
  \href{https://arxiv.org/abs/2303.04168}{arXiv:2303.04168}}.

\bibitem{KamLAND-Zen:2016pfg}
{\bfseries KamLAND-Zen} Collaboration, ``{Search for Majorana Neutrinos near
  the Inverted Mass Hierarchy Region with KamLAND-Zen},''
  \href{https://dx.doi.org/10.1103/PhysRevLett.117.082503}{Phys.\  Rev.\
  Lett.\  {\bfseries 117} (2016) 082503} {\ttfamily
  [\href{https://arxiv.org/abs/1605.02889}{arXiv:1605.02889}]}. [Addendum:
  Phys.Rev.Lett. 117, 109903 (2016)].

\bibitem{GERDA:2018pmc}
{\bfseries GERDA} Collaboration, ``{Improved Limit on Neutrinoless
  Double-$\beta$ Decay of $^{76}$Ge from GERDA Phase II},''
  \href{https://dx.doi.org/10.1103/PhysRevLett.120.132503}{Phys.\  Rev.\
  Lett.\  {\bfseries 120} (2018) 132503} {\ttfamily
  [\href{https://arxiv.org/abs/1803.11100}{arXiv:1803.11100}]}.

\bibitem{Chrzaszcz:2019inj}
M.~Chrzaszcz, {\em et al.}, ``{A frequentist analysis of three right-handed
  neutrinos with GAMBIT},''
  \href{https://dx.doi.org/10.1140/epjc/s10052-020-8073-9}{Eur.\  Phys.\  J.\
  C {\bfseries 80} (2020) 569} {\ttfamily
  [\href{https://arxiv.org/abs/1908.02302}{arXiv:1908.02302}]}.

\bibitem{Mohapatra:1986bd}
R.~N.~Mohapatra and J.~W.~F.~Valle, ``{Neutrino Mass and Baryon Number
  Nonconservation in Superstring Models},''
  \href{https://dx.doi.org/10.1103/PhysRevD.34.1642}{Phys.\  Rev.\  D
  {\bfseries 34} (1986) 1642}.

\bibitem{Mohapatra:1986aw}
R.~N.~Mohapatra, ``{Mechanism for Understanding Small Neutrino Mass in
  Superstring Theories},''
  \href{https://dx.doi.org/10.1103/PhysRevLett.56.561}{Phys.\  Rev.\  Lett.\
  {\bfseries 56} (1986) 561--563}.

\bibitem{Nandi:1985uh}
S.~Nandi and U.~Sarkar, ``{A Solution to the Neutrino Mass Problem in
  Superstring E6 Theory},''
  \href{https://dx.doi.org/10.1103/PhysRevLett.56.564}{Phys.\  Rev.\  Lett.\
  {\bfseries 56} (1986) 564}.

\bibitem{BhupalDev:2012jvh}
P.~S.~Bhupal~Dev and A.~Pilaftsis, ``{Light and Superlight Sterile Neutrinos in
  the Minimal Radiative Inverse Seesaw Model},''
  \href{https://dx.doi.org/10.1103/PhysRevD.87.053007}{Phys.\  Rev.\  D
  {\bfseries 87} (2013) 053007} {\ttfamily
  [\href{https://arxiv.org/abs/1212.3808}{arXiv:1212.3808}]}.

\bibitem{Shrock:1980ct}
R.~E.~Shrock, ``{General Theory of Weak Leptonic and Semileptonic Decays. 1.
  Leptonic Pseudoscalar Meson Decays, with Associated Tests For, and Bounds on,
  Neutrino Masses and Lepton Mixing},''
  \href{https://dx.doi.org/10.1103/PhysRevD.24.1232}{Phys.\  Rev.\  D
  {\bfseries 24} (1981) 1232}.

\bibitem{Bryman:1983cja}
J.~Tran Thanh~Van, ed., ``{Search for Massive Neutrinos in $\pi \to \nu_e$
  Decay},'' \href{https://dx.doi.org/10.1103/PhysRevLett.50.1546}{Phys.\  Rev.\
   Lett.\  {\bfseries 50} (1983) 1546}.

\bibitem{Britton:1992pg}
D.~I.~Britton {\em et~al.}, ``{Measurement of the $\pi^+ \to e^+ \nu$ branching
  ratio},'' \href{https://dx.doi.org/10.1103/PhysRevLett.68.3000}{Phys.\  Rev.\
   Lett.\  {\bfseries 68} (1992) 3000--3003}.

\bibitem{Britton:1992xv}
D.~I.~Britton {\em et~al.}, ``{Improved search for massive neutrinos in $\pi^+
  \to e^+ \nu$ decay},''
  \href{https://dx.doi.org/10.1103/PhysRevD.46.R885}{Phys.\  Rev.\  D
  {\bfseries 46} (1992) R885--R887}.

\bibitem{Abada:2012mc}
A.~Abada, D.~Das, A.~M.~Teixeira, A.~Vicente, and C.~Weiland, ``{Tree-level
  lepton universality violation in the presence of sterile neutrinos: impact
  for $R_K$ and $R_\pi$},''
  \href{https://dx.doi.org/10.1007/JHEP02(2013)048}{JHEP {\bfseries 02} (2013)
  048} {\ttfamily [\href{https://arxiv.org/abs/1211.3052}{arXiv:1211.3052}]}.

\bibitem{Abada:2013aba}
A.~Abada, A.~M.~Teixeira, A.~Vicente, and C.~Weiland, ``{Sterile neutrinos in
  leptonic and semileptonic decays},''
  \href{https://dx.doi.org/10.1007/JHEP02(2014)091}{JHEP {\bfseries 02} (2014)
  091} {\ttfamily [\href{https://arxiv.org/abs/1311.2830}{arXiv:1311.2830}]}.

\bibitem{NA62:2012lny}
{\bfseries NA62} Collaboration, ``{Precision Measurement of the Ratio of the
  Charged Kaon Leptonic Decay Rates},''
  \href{https://dx.doi.org/10.1016/j.physletb.2013.01.037}{Phys.\  Lett.\  B
  {\bfseries 719} (2013) 326--336} {\ttfamily
  [\href{https://arxiv.org/abs/1212.4012}{arXiv:1212.4012}]}.

\bibitem{Derbin:1993wy}
A.~I.~Derbin, {\em et al.}, ``{Experiment on anti-neutrino scattering by
  electrons at a reactor of the Rovno nuclear power plant},'' JETP Lett.\
  {\bfseries 57} (1993) 768--772.
  {\url{http://jetpletters.ru/ps/1202/index.shtml}}.

\bibitem{Hagner:1995bn}
C.~Hagner, {\em et al.}, ``{Experimental search for the neutrino decay $\nu_3
  \to \nu_j + e^+ + e^-$ and limits on neutrino mixing},''
  \href{https://dx.doi.org/10.1103/PhysRevD.52.1343}{Phys.\  Rev.\  D
  {\bfseries 52} (1995) 1343--1352}.

\bibitem{Borexino:2013bot}
{\bfseries Borexino} Collaboration, ``{New limits on heavy sterile neutrino
  mixing in B8 decay obtained with the Borexino detector},''
  \href{https://dx.doi.org/10.1103/PhysRevD.88.072010}{Phys.\  Rev.\  D
  {\bfseries 88} (2013) 072010} {\ttfamily
  [\href{https://arxiv.org/abs/1311.5347}{arXiv:1311.5347}]}.

\bibitem{deGouvea:2019qaz}
A.~de~Gouv\^ea, {\em et al.}, ``{Leptonic Scalars at the LHC},''
  \href{https://dx.doi.org/10.1007/JHEP07(2020)142}{JHEP {\bfseries 07} (2020)
  142} {\ttfamily [\href{https://arxiv.org/abs/1910.01132}{arXiv:1910.01132}]}.

\bibitem{Planck:2018vyg}
{\bfseries Planck} Collaboration, ``{Planck 2018 results. VI. Cosmological
  parameters},'' \href{https://dx.doi.org/10.1051/0004-6361/201833910}{Astron.\
   Astrophys.\  {\bfseries 641} (2020) A6} {\ttfamily
  [\href{https://arxiv.org/abs/1807.06209}{arXiv:1807.06209}]}. [Erratum:
  Astron.Astrophys. 652, C4 (2021)].

\bibitem{Cadamuro:2010cz}
D.~Cadamuro, S.~Hannestad, G.~Raffelt, and J.~Redondo, ``{Cosmological bounds
  on sub-MeV mass axions},''
  \href{https://dx.doi.org/10.1088/1475-7516/2011/02/003}{JCAP {\bfseries 02}
  (2011) 003} {\ttfamily
  [\href{https://arxiv.org/abs/1011.3694}{arXiv:1011.3694}]}.

\bibitem{Dunsky:2022uoq}
D.~I.~Dunsky, L.~J.~Hall, and K.~Harigaya, ``{Dark Radiation Constraints on
  Heavy QCD Axions}.'' {\ttfamily
  \href{https://arxiv.org/abs/2205.11540}{arXiv:2205.11540}}.

\bibitem{Dolan:2017osp}
M.~J.~Dolan, T.~Ferber, C.~Hearty, F.~Kahlhoefer, and K.~Schmidt-Hoberg,
  ``{Revised constraints and Belle II sensitivity for visible and invisible
  axion-like particles},''
  \href{https://dx.doi.org/10.1007/JHEP12(2017)094}{JHEP {\bfseries 12} (2017)
  094} {\ttfamily [\href{https://arxiv.org/abs/1709.00009}{arXiv:1709.00009}]}.
  [Erratum: JHEP 03, 190 (2021)].

\bibitem{Gori:2020xvq}
S.~Gori, G.~Perez, and K.~Tobioka, ``{KOTO vs. NA62 Dark Scalar Searches},''
  \href{https://dx.doi.org/10.1007/JHEP08(2020)110}{JHEP {\bfseries 08} (2020)
  110} {\ttfamily [\href{https://arxiv.org/abs/2005.05170}{arXiv:2005.05170}]}.

\bibitem{Pontecorvo:1957qd}
B.~Pontecorvo, ``{Inverse beta processes and nonconservation of lepton
  charge},'' Zh.\  Eksp.\  Teor.\  Fiz.\  {\bfseries 34} (1957) 247.

\bibitem{Maki:1962mu}
Z.~Maki, M.~Nakagawa, and S.~Sakata, ``{Remarks on the unified model of
  elementary particles},'' \href{https://dx.doi.org/10.1143/PTP.28.870}{Prog.\
  Theor.\  Phys.\  {\bfseries 28} (1962) 870--880}.

\bibitem{Parke:2015goa}
S.~Parke and M.~Ross-Lonergan, ``{Unitarity and the three flavor neutrino
  mixing matrix},'' \href{https://dx.doi.org/10.1103/PhysRevD.93.113009}{Phys.\
   Rev.\  D {\bfseries 93} (2016) 113009} {\ttfamily
  [\href{https://arxiv.org/abs/1508.05095}{arXiv:1508.05095}]}.

\bibitem{Fong:2016yyh}
C.~S.~Fong, H.~Minakata, and H.~Nunokawa, ``{A framework for testing leptonic
  unitarity by neutrino oscillation experiments},''
  \href{https://dx.doi.org/10.1007/JHEP02(2017)114}{JHEP {\bfseries 02} (2017)
  114} {\ttfamily [\href{https://arxiv.org/abs/1609.08623}{arXiv:1609.08623}]}.

\bibitem{Ellis:2020hus}
S.~A.~R.~Ellis, K.~J.~Kelly, and S.~W.~Li, ``{Current and Future Neutrino
  Oscillation Constraints on Leptonic Unitarity},''
  \href{https://dx.doi.org/10.1007/JHEP12(2020)068}{JHEP {\bfseries 12} (2020)
  068} {\ttfamily [\href{https://arxiv.org/abs/2008.01088}{arXiv:2008.01088}]}.

\bibitem{Hu:2020oba}
Z.~Hu, J.~Ling, J.~Tang, and T.~Wang, ``{Global oscillation data analysis on
  the $3\nu$ mixing without unitarity},''
  \href{https://dx.doi.org/10.1007/JHEP01(2021)124}{JHEP {\bfseries 01} (2021)
  124} {\ttfamily [\href{https://arxiv.org/abs/2008.09730}{arXiv:2008.09730}]}.

\bibitem{Goldhagen:2021kxe}
K.~Goldhagen, M.~Maltoni, S.~E.~Reichard, and T.~Schwetz, ``{Testing sterile
  neutrino mixing with present and future solar neutrino data},''
  \href{https://dx.doi.org/10.1140/epjc/s10052-022-10052-2}{Eur.\  Phys.\  J.\
  C {\bfseries 82} (2022) 116} {\ttfamily
  [\href{https://arxiv.org/abs/2109.14898}{arXiv:2109.14898}]}.

\bibitem{Ma:2023kfr}
P.-X.~Ma, {\em et al.}, ``{Lattice QCD Calculation of Electroweak Box
  Contributions to Superallowed Nuclear and Neutron Beta Decays}.'' {\ttfamily
  \href{https://arxiv.org/abs/2308.16755}{arXiv:2308.16755}}.

\end{thebibliography}\endgroup

\end{document}